# Supersymmetric Jaynes-Cummings model and its exact solutions

A. D. Alhaidari

*Shura Council, Riyadh 11212, Saudi Arabia*
E-mail: haidari@mailaps.org

The super-algebraic structure of a generalized version of the Jaynes-Cummings model is investigated. We find that a $Z_2$ graded extension of the so(2,1) Lie algebra is the underlying symmetry of this model. It is isomorphic to the four-dimensional super-algebra u(1/1) with two odd and two even elements. Differential matrix operators are taken as realization of the elements of the superalgebra to which the model Hamiltonian belongs. Several examples with various choices of superpotentials are presented. The energy spectrum and corresponding wavefunctions are obtained analytically.

## I. INTRODUCTION

The Jaynes-Cummings (JC) model is a simple model for the description of one-photon exchange with a two-level atomic system [1]. The model has been the target of extensive studies in the literature [2]. Its simplicity allows exact analytic application of the fundamental laws of quantum electrodynamics. Moreover, its exact solvability in the rotating wave approximation exhibits interesting quantum mechanical effects like the collapses and revivals of Rabi oscillations [3]. These effects have important applications in optical communication [4] and laser trapping and cooling of atoms [5]. It enables one to study, in a simple but realistic way, not only the coherent properties of the quantized field, but also its influence on atoms (e.g., collapses and revivals [6], squeezing [7], the interaction of a trapped ion with a laser field [8], etc.). The JC model has been the subject of many generalizations including multi-level and multi-mode systems [9], intensity-dependent and time-dependent coupling [10]. The JC model was generalized using the methods of supersymmetric quantum mechanics where the usual creation and annihilation operators of the harmonic oscillator become generalized creation and annihilation operators satisfying the supersymmetric algebra [11]. Applications of the supersymmetric approach to the JC model, including representation theory of super-algebras, have opened large venues for the exact solvability of the model. In fact, as will be demonstrated here, the supersymmetric JC Hamiltonian is an element of the u(1/1) superalgebra [12]. Additionally, shape invariance studies of bound state problems for two-level systems also lead to generalized JC models [11]. In the case of exact resonance, the u(1/1) dynamical superalgebra reduces to $N = 2$ supersymmetric algebra where the JC model coincides with the supersymmetric harmonic oscillator (the standard JC model).

In the atomic units, the Hamiltonian of the standard JC model is written in terms of creation and annihilation operators, $a^\dagger$ and $a$, as

$$H_{JC} = \mu \sigma_3 + \lambda \left( a \sigma_+ + a^\dagger \sigma_- \right) + a^\dagger a, \tag{1}$$

where $\mu$ is the frequency of the model (depending on the physical interpretation of the model, it could also be proportional to the atomic transition frequency or detuning frequency) and $\lambda$ is the coupling parameter. $\{\sigma_i\}$ are the three Pauli matrices

$$\sigma_3 = \begin{pmatrix} 1 & 0 \\ 0 & -1 \end{pmatrix},\ \sigma_+ = \begin{pmatrix} 0 & 1 \\ 0 & 0 \end{pmatrix},\ \sigma_- = \begin{pmatrix} 0 & 0 \\ 1 & 0 \end{pmatrix}. \tag{2}$$

Using the usual realization of creation and annihilation operators, we write

$$a^\dagger = \frac{d}{dx} + \rho x,\ a = -\frac{d}{dx} + \rho x, \tag{3}$$

where $\rho$ is proportional to the square of the mode frequency. Therefore, we can write the JC Hamiltonian (1) explicitly as follows

$$H_{JC} = \begin{pmatrix} \mu + \rho & \lambda\left(\rho x - \frac{d}{dx}\right) \\ \lambda\left(\rho x + \frac{d}{dx}\right) & -\mu + \rho \end{pmatrix} - \frac{d^2}{dx^2} + \rho^2 x^2. \tag{1'}$$

In this work, we exploit the (super-) symmetry of a generalized version of the JC model and obtain some of its relevant and interesting analytic solutions. We show that this generalized model carries a representation of a superalgebra which is a $Z_2$ graded extension of the so(2,1) Lie algebra. This superalgebra was introduced in [13] while searching for solutions of the Dirac equation where we found that the "canonical form" of the Dirac Hamiltonian is an element of this superalgebra. It is the lowest in a special superalgebra hierarchy [14]. Moreover, it is isomorphic to u(1/1) Lie superalgebra [12]. Most of earlier work on the analytic solutions of the two-level JC model was limited to the underlying symmetries associated with $N = 2$ supersymmetry, su(2) or su(1,1) symmetry. Moreover, the superpotentials used in the supersymmetric studies were very limited. Here, as well, we will not be able to exhaust all possibilities but give several exact solutions for interesting and non-trivial superpotential examples. Now, so(2,1) algebra is a three dimensional Lie algebra with basis elements satisfying the commutation relations $[L_3, L_\pm] = \pm L_\pm$ and $[L_+, L_-] = -L_3$. It is very useful and highly important in various physical applications, and in the solution of many three-parameter problems. It has been studied extensively in the literature as a potential algebra and spectrum-generating algebra for several problems [15]. Following Kac and others [16], we define a superalgebra $\mathcal{G}$ as the $Z_2$ graded algebra $\mathcal{G} = \mathcal{G}_0 + \mathcal{G}_1$ with a product operation $\circ$ satisfying $p \circ q = -(-1)^{\sigma(p,q)} q \circ p$, where $\sigma(p,q) = \deg(p) \times \deg(q)$ and $\deg(p) = m \leftrightarrow p \in \mathcal{G}_m$, $m = 0$ or 1. An element of $\mathcal{G}$ is called even (odd) if it belongs to $\mathcal{G}_0$ ($\mathcal{G}_1$). We call the anti-symmetric product operation $\circ$ which involves an even element the commutator and designate it by [ , ] while the symmetric operation that involves only odd elements is called the anti-commutator, and is designated by { , }. The $Z_2$ grading of so(2,1) Lie algebra of concern to us is a four-dimensional super-algebra, with two odd elements $L_\pm$ and two even elements, $L_0$ and $L_3$, satisfying the commutation/anticommutation relations [13]:

$$[L_3, L_\pm] = \pm L_\pm,\ \{L_+, L_-\} = L_0,\ [L_0, L_3] = [L_0, L_\pm] = 0, \tag{4}$$

where $L_\pm^\dagger = L_\mp$, which implies hermiticity of the even operators (i.e., $L_0^\dagger = L_0$ and $L_3^\dagger = L_3$). These relations also show that $L_0$ belongs to the center of the superalgebra since it commutes with all of its elements. Additionally, this superalgebra has a second order Casimir invariant operator, which could be written as

$$C_2 = L_0 L_3 + L_- L_+. \tag{5}$$

This algebra is, in fact, isomprhic to the u(1/1) superalgebra [12]. In the present settings, we are not concerened with the general representations of this superalgebra. We will only be interested in a special realization of its elements. These are 2×2 matrices of differential operators acting in a two-component $L^2$ function space with elements $\chi(x) = \begin{pmatrix} \phi_+ \\ \phi_- \end{pmatrix}$:

$$L_\pm = \sigma_\pm \left[ W(x) \mp \tfrac{d}{dx} \right], \quad L_3 = \tfrac{1}{2}\sigma_3, \tag{6}$$

where $W(x)$ is a real differentiable function (the "superpotential"). Using the anti-commutation relation in (4) we obtain:

$$L_0 = \begin{pmatrix} -\tfrac{d^2}{dx^2} + W^2 - W' & 0 \\ 0 & -\tfrac{d^2}{dx^2} + W^2 + W' \end{pmatrix}, \tag{7}$$

where $W' = dW/dx$. In this representation, the Casimir operator (5) is not independent but, in fact, we obtain $C_2 = \tfrac{1}{2} L_0$. The odd operators $L_\pm$ are the raising and lowering (creation and annihilation) operators in a two-component Hilbert space. They are first order in the derivatives, whereas the even operators are zero and second order. $L_3$ is the helicity/parity operator. Now, since $L_0$ is in the center of the algebra, then its eigenvalue is an invariant and we can write $L_0 |\chi\rangle \sim |\chi\rangle$. Therefore, one can interpret the diagonal elements of $L_0$ as Schrödinger operators resulting in the identification of the two potential functions $W^2 \pm W'$. In supersymmetric quantum mechanics these are the two isospectral (i.e., having the same spectrum) superpartner potentials [17]. If a linear operator $H$ belongs to the superalgebra (i.e., carries a representation of this super-symmetry), then $H$ could be expanded as a linear combination of these four basis elements as follows:

$$H = \alpha_3 L_3 + \alpha_+ L_+ + \alpha_- L_- + \alpha_0 L_0, \tag{8}$$

where the $\alpha$'s are constant parameters. Requiring that this operator be hermitian gives $\alpha_\pm^* = \alpha_\mp$, $\alpha_0^* = \alpha_0$, and $\alpha_3^* = \alpha_3$, which yields:

$$H = \begin{pmatrix} \mu & \lambda\left(W - \tfrac{d}{dx}\right) \\ \lambda\left(W + \tfrac{d}{dx}\right) & -\mu \end{pmatrix} + \begin{pmatrix} -\tfrac{d^2}{dx^2} + W^2 - W' & 0 \\ 0 & -\tfrac{d^2}{dx^2} + W^2 + W' \end{pmatrix}, \tag{9}$$

where $\lambda = \alpha_\pm$, $\mu = 2\alpha_3$ and $\alpha_0 = 1$. Since $L_0$ commutes with $H$, then a two-component representation, $\chi$, could be found such that $H|\chi\rangle \sim |\chi\rangle$ and $L_0|\chi\rangle \sim |\chi\rangle$. Requiring that $H$ be linear in the derivatives makes $\alpha_0 = 0$ and results in $H$ being identical to the Dirac Hamiltonian (in the relativistic units $\hbar = c = 1$) for a spinor of mass $\mu$ coupled to the pseudo-scalar potential $W(x)$ and with $\lambda = 1$ [13]. This case will be discussed briefly below.

The two-parameter Hamiltonian in (8) or (9) could be considered as a generalized version of the JC Hamiltonian (1) which is rewritten as

$$H = \mu \sigma_3 + \lambda \left( A\sigma_+ + A^\dagger \sigma_- \right) + A^\dagger \sigma_- A \sigma_+ + A \sigma_+ A^\dagger \sigma_- \tag{10}$$

One of the two sources of generalizations in the model is the replacement of the linear function $\rho x$ in the creation and annihilation operators by $W(x)$. That is, $A^\dagger = \tfrac{d}{dx} + W(x)$ and $A = -\frac{d}{dx} + W(x)$ [1]. In other words, the oscillating electromagnetic wave is replaced by a general bosonic field depicted by the superpotential $W(x)$. Moreover, since $A^\dagger A = -\frac{d^2}{dx^2} + W^2 + W'$ and $A A^\dagger = -\frac{d^2}{dx^2} + W^2 - W'$, then neither of the two superpartner potentials $W^2 \pm W'$ should assume precedence over the other. Consequently, $A A^\dagger$ as well as $A^\dagger A$ are included in the Hamiltonian, which is the source of the second generalization in the model. Variations of this two-level generalized JC model have already been treated in the literature [See, for example, Refs. 9-12]. However, analytic solutions were obtained for only a very limited selection of superpotentials, $W(x)$. Now, we choose a two-component representation $\chi = \begin{pmatrix} \phi^+ \\ \phi^- \end{pmatrix}$ parameterized by the two real eigenvalues of $L_0$ and $H$ as follows

$$L_0 |\chi\rangle = \omega |\chi\rangle, \text{ and } H |\chi\rangle = \varepsilon |\chi\rangle. \tag{11}$$

Using the explicit realization of these operators in Eq. (7) and Eq. (9) we obtain

$$\left(-\frac{d^2}{dx^2} + W^2 \mp W'\right)\phi^\pm(x) = \omega \phi^\pm(x), \tag{12}$$

$$\begin{pmatrix} \mu + \omega & \lambda\left(W - \frac{d}{dx}\right) \\ \lambda\left(W + \frac{d}{dx}\right) & -\mu + \omega \end{pmatrix} \begin{pmatrix} \phi^+ \\ \phi^- \end{pmatrix} = \varepsilon \begin{pmatrix} \phi^+ \\ \phi^- \end{pmatrix}. \tag{13}$$

Equation (13) gives, and is equivalent to, the "kinetic balance" relation

$$\phi^\mp = \frac{\lambda}{\varepsilon - \omega \pm \mu}\left(W \pm \frac{d}{dx}\right)\phi^\pm, \tag{13'}$$

where $\varepsilon \neq \omega \mp \mu$, respectively. Compatibility of Eq. (12) and (13)' results in the following relation between the two eigenvalues $\omega$ and $\varepsilon$

$$\varepsilon = \omega \pm \sqrt{\mu^2 + \lambda^2 \omega}, \tag{14}$$

where $\omega \geq -(\mu/\lambda)^2$ is required for the reality of the representation.

There is an interesting relation between this model and the Dirac Hamiltonian. The Author and co-workers are currently investigating this relation [18]. It might be instructive to give a very brief outline of this relation as follows. In the relativistic units ($\hbar = c = 1$) the one-dimensional Dirac Hamiltonian with coupling to vector, scalar, and pseudo-scalar potentials reads as follows

$$H_D = \begin{pmatrix} m + V + S & -\frac{d}{dx} + W \\ \frac{d}{dx} + W & -m + V - S \end{pmatrix}, \tag{15}$$

where $V(x)$ is the vector potential, $S(x)$ the scalar potential, and $W(x)$ is the pseudo-scalar[2]. Taking $V = S = 0$ and the two-component spinor as $\psi = \begin{pmatrix} \phi^+ \\ \phi^- \end{pmatrix}$, then the Dirac equation $(H_D - E)\psi = 0$ gives $\phi^\mp = \frac{1}{E \pm m}\left(W \pm \frac{d}{dx}\right)\phi^\pm$ and

$$\begin{pmatrix} -\frac{d^2}{dx^2}+W^2-W' & 0 \\ 0 & -\frac{d^2}{dx^2}+W^2+W' \end{pmatrix} \begin{pmatrix} \phi^+ \\ \phi^- \end{pmatrix} = (E^2-m^2)\begin{pmatrix} \phi^+ \\ \phi^- \end{pmatrix}, \qquad (16)$$

where $E$ is the relativistic energy. Thus, the Dirac Hamiltonian (15) with $V = S = 0$ is equivalent to the generalized JC Hamiltonian (9) with $\mu = m$ and $\lambda = 1$. This is so, because the difference between them ($H - H_D$) in the solution space is only a constant, $E^2 - m^2$.

---

[1] One could think of these as conventional creation and annihilation operators in a new configuration space with coordinate $y(x)$ (i.e., $\pm \frac{d}{dy} + \rho y$) with $W(x) = \rho y (dy/dx)$.

[2] If $W(x)$ could be written as $\frac{\kappa}{x} + U(x)$ with $\kappa = \pm 1, \pm 2, ...$, then $H_D$ could also be interpreted as the Dirac Hamiltonian in three dimensions with spherical symmetry where $x \in [0, \infty]$ is the radial coordinate, $\kappa$ stands for the spin-orbit quantum number, and $U(x)$ is the radial pseudo scalar potential.

Next, we give several nonrelativistic examples of the model (10). The original JC model where $W(x)$ is linear in $x$, which has been studied extensively in the literature, will not be discussed.

## II. EXAMPLES

In this section, we present several examples for the generalized model (10) each of which is associated with a given choice of the "superpotential" function $W(x)$. The energy spectrum of the bound states and the corresponding two-component wave function are obtained analytically. Our method of solution goes as follows. A coordinate trans-formation $y(x)$ to a new configuration space $y \in [y_-, y_+] \subseteq \Re$ is carried out. In the new space, we propose the following two components of the wavefunction

$$\phi^\pm(x) = A^\pm \Omega_\pm(y) f_\pm(y), \qquad (17)$$

where $A^\pm$ are the normalization constants, $\Omega_\pm(y)$ is the weight function that forces compatibility with the boundary conditions, and $f_\pm(y)$ is chosen to be the hypergeometric or confluent hypergeometric series. We require that the series terminates so that the wavefunction is normalizable by being square integrable (i.e., the integral $\int_{y_-}^{y_+} \frac{1}{y'}[\Omega_\pm(y)f_\pm(y)]^2 dy$ is finite). Substituting (17) into (12) gives

$$\left[-\frac{d^2}{dy^2} - \left(\frac{y''}{y'^2} + 2\frac{\Omega_\pm'}{\Omega_\pm}\right)\frac{d}{dy} - \frac{\Omega_\pm''}{\Omega_\pm} - \frac{y''}{y'^2}\left(\frac{\Omega_\pm'}{\Omega_\pm}\right) + \frac{W^2}{y'^2} \mp \frac{W'}{y'} - \frac{\omega}{y'^2}\right] f_\pm(y) = 0, \qquad (18)$$

where the primes on $y$ stand for the derivative with respect to $x$ whereas other primes (on $\Omega_\pm$ and $W$) mean the derivative with respect to $y$. It was assumed that $\Omega_\pm(y)$ is regular on the interval $[y_-, y_+]$. Equation (13) or its equivalent, the kinetic balance relation (13)', determines the relative normalization of the two components of the wave function, $A^+/A^-$. Now, we will be considering transformations to two types of configuration spaces: a bounded one where $y \in [-1,+1]$, and a semi-infinite one where $y \in [0,\infty]$. The weight function for the former is $\Omega_\pm(y) = (1-y)^{\alpha_\pm}(1+y)^{\beta_\pm}$ whereas for the latter it is taken as $\Omega_\pm(y) = y^{\alpha_\pm} e^{-\beta_\pm y}$, where appropriate conditions are imposed on the real parameters $\alpha_\pm$ and $\beta_\pm$ (mostly, being positive). These weight functions force the wavefunction to vanish at the boundary since $f_\pm(y)$ is regular there. Requiring that $f_\pm(y)$ be hypergeometric in the former case dictates that the term $y''/y'^2$ multiplying the first order derivative in Eq. (18) be the ratio of a linear function in $y$ divided by $1-y^2$. This is satisfied by taking $y(x)$ to be a hyperbolic function (e.g., $\tanh \tau x$, $\cosh \tau x$, $\mathrm{sech}\, \tau x$, etc.). On the other hand, for the latter case where $f_\pm(y)$ is the confluent hypergeometric function, $y''/y'^2$ must be the sum of a constant and a term proportional to $y^{-1}$. This could be accomplished by taking $y$ to be a monomial $y^\tau$, an exponential $e^{-\tau x}$, or logarithmic $\ln \tau x$. To select the appropriate superpotential $W(y)$ and to fix the real parameter $\tau$ in $y(x)$ we require that the terms in Eq. (18) without derivatives to be the same as those in the differential equation of the (confluent-) hypergeometric function, respectively. Putting all elements of this strategy for the analytic solution together, we end up with few possibilities some of which are chosen as examples in the following subsections.

### A. Inverse linear superpotential

In this case: $W(x) = \kappa x^{-1} + \gamma$, $y(x) = \tau x$, $x \in [0, \infty]$, and $\Omega_\pm(y) = y^{\alpha_\pm} e^{-\beta_\pm y}$, where $\kappa$ and $\gamma$ are real and $\tau$ positive. Therefore, the physical parameters of the model are $\{\mu, \lambda, \kappa, \gamma\}$. The resulting differential equation (18) becomes

$$\left\{ -\frac{d^2}{dy^2} - 2\left(\frac{\alpha_\pm}{y} - \beta_\pm\right)\frac{d}{dy} + \frac{1}{y^2}\left[\left(\kappa \pm \tfrac{1}{2}\right)^2 - \left(\alpha_\pm - \tfrac{1}{2}\right)^2\right] \right. $$
$$\left. + 2\frac{\alpha_\pm \beta_\pm + \kappa\gamma/\tau}{y} - \beta_\pm^2 + \frac{\gamma^2 - \omega}{\tau^2} \right\} f_\pm(y) = 0 \qquad (19)$$

Comparing this with the differential equation of the confluent hypergeometric series ${}_1F_1(a_\pm; c_\pm; y)$ [19], we obtain

$$\beta_\pm = \tfrac{1}{2}, \quad \tau = 2\sqrt{\gamma^2 - \omega}, \quad \alpha_\pm = |\kappa| + \tfrac{1}{2}\left(1 \pm \tfrac{\kappa}{|\kappa|}\right) = \begin{cases} \kappa + (1\pm 1)/2 & ,\kappa > 0 \\ -\kappa + (1\mp 1)/2 & ,\kappa < 0 \end{cases}, \qquad (20a)$$

$$c_\pm = 2\alpha_\pm, \quad a_\pm = \alpha_\pm + 2\kappa\gamma/\tau. \qquad (20b)$$

Reality of the solution puts a stronger constraint on $\omega$, the eigenvalues of $L_0$, which is that $\gamma^2 > \omega \geq -(\mu/\lambda)^2$. For bound states, square integrability requires that the confluent hypergeometric series ${}_1F_1(a_\pm; c_\pm; y)$ terminate. This is accomplished by choosing $a_\pm$ to be a negative integer or zero [19]. Imposing this requirement on the solutions in (20) restricts $\omega$ to be an element of the following infinite discrete set,

$$\omega_n = \gamma^2\left[1 - \left(\frac{\kappa}{n+|\kappa|}\right)^2\right], \qquad (21)$$

where $n = 0, 1, 2, \ldots$ Moreover, it also dictates that bound state solutions are possible only if the physical parameters $\kappa$ and $\gamma$ are of opposite sign (i.e., $\gamma\kappa < 0$). These discrete values of $\omega$ conform to the requirement of reality of the representation, which is that $\gamma^2 > \omega_n \geq -(\mu/\lambda)^2$. In fact, Eq. (21) shows that $\gamma^2 > \omega_n \geq 0$. Now, for a given $n$ (i.e., for a given $\omega_n$), we obtain $a_\pm = -n + \tfrac{1}{2}(1 \pm \kappa/|\kappa|)$. Inserting $\omega_n$ for $\omega$ in Eq. (14) gives the bound states energy spectrum $\{\varepsilon_n\}_{n=0}^\infty$. The corresponding components of the wave function are

$$\phi_n^+(x) = A_n^+ e^{-\tau_n x/2} \begin{cases} (\tau_n x)^{\kappa+1} {}_1F_1(-n+1; 2\kappa+2; \tau_n x) & ,\kappa > 0 \\ (\tau_n x)^{-\kappa} {}_1F_1(-n; -2\kappa; \tau_n x) & ,\kappa < 0 \end{cases} \qquad (22a)$$

$$\phi_n^-(x) = A_n^- e^{-\tau_n x/2} \begin{cases} (\tau_n x)^\kappa {}_1F_1(-n; 2\kappa; \tau_n x) & , \kappa > 0 \\ (\tau_n x)^{-\kappa+1} {}_1F_1(-n+1; -2\kappa+2; \tau_n x) & , \kappa < 0 \end{cases} \quad (22b)$$

where $\tau_n = 2\sqrt{\gamma^2 - \omega_n} = 2|\gamma\kappa|/(n+|\kappa|)$ and $n = 1, 2, 3, \ldots$ Ground state is the lowest energy state, which is obtained from Eqs. (22) by setting $n = 0$ (i.e., $\omega = 0$ and $\varepsilon = \pm|\mu|$) and requiring normalizability:

$$\chi_0(x) = A_0^- e^{\gamma x} (-2\gamma x)^\kappa \begin{pmatrix} 0 \\ 1 \end{pmatrix}, \quad \kappa > 0 \quad (23a)$$

$$\chi_0(x) = A_0^+ e^{-\gamma x} (2\gamma x)^{-\kappa} \begin{pmatrix} 1 \\ 0 \end{pmatrix}, \quad \kappa < 0 \quad (23b)$$

### B. Exponential superpotential

In this case: $W(x) = \kappa e^{-\tau x} + \gamma$, $y(x) = e^{-\tau x}$, $x \in [-\infty, +\infty]$, and $\Omega_\pm(y) = y^{\alpha_\pm} e^{-\beta_\pm y}$, where $\kappa$ and $\gamma$ are real and $\tau$ positive. The physical parameters of the model are $\{\mu, \lambda, \kappa, \tau, \gamma\}$ and the differential equation (18) becomes

$$\left\{ -\frac{d^2}{dy^2} - \left(\frac{2\alpha_\pm + 1}{y} - 2\beta_\pm\right)\frac{d}{dy} - \frac{\alpha_\pm^2 + (\omega - \gamma^2)/\tau^2}{y^2} \right.$$
$$\left. + \frac{\beta_\pm(2\alpha_\pm + 1) + (2\kappa\gamma/\tau^2) \pm (\kappa/\tau)}{y} - \beta_\pm^2 + \frac{\kappa^2}{\tau^2} \right\} f_\pm(y) = 0 \quad (24)$$

Comparing this with the differential equation of the confluent-hypergeometric series ${}_1F_1(a_\pm; c_\pm; y)$, we obtain

$$\beta_\pm = \tfrac{1}{2}, \quad \tau = 2|\kappa|, \quad \alpha_\pm = \tfrac{1}{2|\kappa|}\sqrt{\gamma^2 - \omega} \equiv \alpha, \quad (25a)$$

$$c_\pm = 2\alpha + 1, \quad a_\pm = \alpha + \tfrac{1}{2}\tfrac{\gamma}{\kappa} + \tfrac{1}{2}\left(1 \pm \tfrac{\kappa}{|\kappa|}\right). \quad (25b)$$

Thus, the physical parameters are reduced by one to $\{\mu, \lambda, \kappa, \gamma\}$. Reality of the representation requires that $\gamma^2 > \omega \geq -(\mu/\lambda)^2$. Moreover, it is easy to note that $a_\pm = a_\mp + 1$ for $\pm\kappa > 0$. Again, to obtain normalizable bound states the confluent hypergeometric series must terminate. This requires that $a_\pm$ be a negative integer or zero and that bound state solutions are possible only if the physical parameters $\kappa$ and $\gamma$ are of opposite signs (i.e., $\gamma\kappa < 0$). Consequently, the eigenvalue $\omega$ becomes an element of the following discrete set

$$\omega_n = \gamma^2 - 4\kappa^2\left(n - \tfrac{1}{2}|\gamma/\kappa|\right)^2, \quad (26)$$

where $n = 0, 1, \ldots, N$ and $N$ is the largest integer that satisfies $N \leq \tfrac{1}{2}|\gamma/\kappa|$. Thus, the reality requirement of the representation, $\gamma^2 > \omega_n \geq -(\mu/\lambda)^2$, is automatically satisfied since, in fact, $\gamma^2 > \omega_n \geq 0$. Substituting $\omega_n$ for $\omega$ in Eq. (14) gives the finite bound states energy spectrum $\{\varepsilon_n\}_{n=0}^N$. The corresponding two components of the wave function are

$$\phi_n^+(x) = A_n^+ \exp\left(-\tfrac{1}{2}e^{-2|\kappa|x}\right)e^{-2|\kappa|\alpha_n x}\begin{cases}{}_1F_1\left(-n+1;2\alpha_n+1;e^{-2\kappa x}\right) &,\kappa>0\\ {}_1F_1\left(-n;2\alpha_n+1;e^{2\kappa x}\right) &,\kappa<0\end{cases} \quad (27a)$$

$$\phi_n^-(x) = A_n^- \exp\left(-\tfrac{1}{2}e^{-2|\kappa|x}\right)e^{-2|\kappa|\alpha_n x}\begin{cases}{}_1F_1\left(-n;2\alpha_n+1;e^{-2\kappa x}\right) &,\kappa>0\\ {}_1F_1\left(-n+1;2\alpha_n+1;e^{2\kappa x}\right) &,\kappa<0\end{cases} \quad (27b)$$

where $\alpha_n = \tfrac{1}{2|\kappa|}\sqrt{\gamma^2-\omega_n} = \tfrac{1}{2}\left|\tfrac{\gamma}{\kappa}\right|-n$ and $n=1,2,..,N$. The lowest energy state is obtained from Eqs. (27) by setting $n=0$ (i.e., $\omega=0$ and $\varepsilon=\pm|\mu|$) and requiring normalizability. It reads as follows:

$$\chi_0(x) = A_0^- \exp\left(-\tfrac{1}{2}e^{-2\kappa x}\right)e^{\gamma x}\begin{pmatrix}0\\1\end{pmatrix}, \quad \kappa>0 \quad (28a)$$

$$\chi_0(x) = A_0^+ \exp\left(-\tfrac{1}{2}e^{2\kappa x}\right)e^{-\gamma x}\begin{pmatrix}1\\0\end{pmatrix}, \quad \kappa<0 \quad (28b)$$

### C. Hyperbolic superpotentials

For this case we take: $W(x)=\kappa\tanh(\tau x)+\gamma$, $y(x)=\tanh(\tau x)$, $x\in[-\infty,+\infty]$, and $\Omega_\pm(y) = (1-y)^{\alpha_\pm}(1+y)^{\beta_\pm}$, where $\kappa$ and $\gamma$ are real and $\tau$ positive. Therefore, the physical parameters of the model are $\{\mu,\lambda,\kappa,\tau,\gamma\}$ and the differential equation (18) becomes

$$\left\{(1-y^2)\frac{d^2}{dy^2} - 2\left[\alpha_\pm - \beta_\pm + y(\alpha_\pm+\beta_\pm+1)\right]\frac{d}{dy} - \left(\beta_\pm^2-\alpha_\pm^2+\kappa\gamma/\tau^2\right)\frac{2y}{1-y^2}\right.$$
$$\left. +2\frac{\alpha_\pm^2+\beta_\pm^2-(\kappa^2+\gamma^2-\omega)/2\tau^2}{1-y^2} - \left(\alpha_\pm+\beta_\pm+\tfrac{1}{2}\right)^2 + \left(\tfrac{\kappa}{\tau}\pm\tfrac{1}{2}\right)^2\right\}f_\pm(y)=0. \quad (29)$$

Comparing this with the differential equation of the hypergeometric series ${}_2F_1\left(a_\pm,b_\pm;c_\pm;\tfrac{1+y}{2}\right)$ [19], we obtain

$$\alpha_\pm = \tfrac{1}{2\tau}\sqrt{(\kappa+\gamma)^2-\omega} \equiv \alpha, \quad \beta_\pm = \tfrac{1}{2\tau}\sqrt{(\kappa-\gamma)^2-\omega} \equiv \beta, \quad (30a)$$

$$c_\pm = 2\beta+1, \quad \left(\alpha+\beta+\tfrac{1}{2}-a_\pm\right)^2 = \left(\tfrac{\kappa}{\tau}\pm\tfrac{1}{2}\right)^2, \quad b_\pm = 2\left(\alpha+\beta+\tfrac{1}{2}\right)-a_\pm. \quad (30b)$$

To simplify the solution (without too much loss of generality) we choose $\gamma=0$. Thus, the physical parameters of the model are reduced to $\{\mu,\lambda,\kappa,\tau\}$ and Eq. (30) gives

$$\alpha=\beta=\tfrac{1}{2\tau}\sqrt{\kappa^2-\omega}, \quad c_\pm = 2\alpha+1, \quad (30a)'$$

$$\left(2\alpha+\tfrac{1}{2}-a_\pm\right)^2 = \left(\tfrac{\kappa}{\tau}\pm\tfrac{1}{2}\right)^2, \quad b_\pm = 4\alpha+1-a_\pm. \quad (30b)'$$

Reality of the representation requires that $\kappa^2 > \omega \geq -(\mu/\lambda)^2$. To obtain normalizable bound states the hypergeometric series ${}_2F_1\left(a_\pm,b_\pm;c_\pm;\tfrac{1+y}{2}\right)$ must terminate. This requires that either $a_\pm$ or $b_\pm$ be a negative integer or zero [19]. Due to the exchange symmetry $a_\pm \leftrightarrow b_\pm$ in $_2F_1\left(a_\pm, b_\pm; c_\pm; \frac{1+y}{2}\right)$ we choose $a_\pm$ to meet this requirement. Consequently, $\omega$ assumes the following discrete values

$$\omega_n = \kappa^2 - \tau^2\left(n - |\kappa|/\tau\right)^2, \tag{31}$$

where $n = 0, 1, .., N$ and $N$ is the largest integer that is less than or equal to $|\kappa|/\tau$. Thus, the reality requirement of the solution is satisfied since $\kappa^2 > \omega_n \geq 0$. Moreover, we obtain $a_- = a_+ + \frac{\kappa}{|\kappa|}$ and the following parameters of the hypergeometric series

$$a_\pm = -n + \tfrac{1}{2}\left(1 \mp \tfrac{\kappa}{|\kappa|}\right),\ b_\pm = a_\mp + 2|\kappa|/\tau,\ c_\pm = -n + 1 + |\kappa|/\tau. \tag{32}$$

Substituting $\omega_n$ from Eq. (31) for $\omega$ in Eq. (14) gives the finite bound states energy spectrum $\{\varepsilon_n\}_{n=0}^N$. The corresponding two components of the wave function are

$$\phi_n^+(x) = A_n^+ \left(\cosh \tau x\right)^{n - |\kappa|/\tau} \begin{cases} _2F_1\left(-n, 2\tfrac{\kappa}{\tau} - n + 1; \tfrac{\kappa}{\tau} - n + 1; \tfrac{1}{2} + \tfrac{1}{2}\tanh \tau x\right) & ,\kappa > 0 \\ _2F_1\left(-n + 1, -2\tfrac{\kappa}{\tau} - n; -\tfrac{\kappa}{\tau} - n + 1; \tfrac{1}{2} + \tfrac{1}{2}\tanh \tau x\right) & ,\kappa < 0 \end{cases} \tag{33a}$$

$$\phi_n^-(x) = A_n^- \left(\cosh \tau x\right)^{n - |\kappa|/\tau} \begin{cases} _2F_1\left(-n + 1, 2\tfrac{\kappa}{\tau} - n; \tfrac{\kappa}{\tau} - n + 1; \tfrac{1}{2} + \tfrac{1}{2}\tanh \tau x\right) & ,\kappa > 0 \\ _2F_1\left(-n, -2\tfrac{\kappa}{\tau} - n + 1; -\tfrac{\kappa}{\tau} - n + 1; \tfrac{1}{2} + \tfrac{1}{2}\tanh \tau x\right) & ,\kappa < 0 \end{cases} \tag{33b}$$

where $n = 1, 2, .., N$. The lowest energy state is obtained from Eqs. (33) by setting $n = 0$ (i.e., $\omega = 0$ and $\varepsilon = \pm|\mu|$) and requiring normalizability giving

$$\chi_0(x) = A_0^+ \left(\cosh \tau x\right)^{-\kappa/\tau} \begin{pmatrix} 1 \\ 0 \end{pmatrix},\ \kappa > 0 \tag{34a}$$

$$\chi_0(x) = A_0^- \left(\cosh \tau x\right)^{\kappa/\tau} \begin{pmatrix} 0 \\ 1 \end{pmatrix},\ \kappa < 0 \tag{34b}$$

Finally, there is an interesting and highly nontrivial example that belongs to this class of superpotentials. This is when $W(x) = \kappa \coth(\tau x) + \gamma \operatorname{csch}(\tau x)$, where $\kappa$ and $\gamma$ are real and $\tau$ positive. Configuration space is the semi-infinite real line, $x \in [0, \infty]$, and the weight function is $\Omega_\pm(y) = (y - 1)^{\alpha_\pm}(y + 1)^{-\beta_\pm}$, where $y = \cosh(\tau x)$ and $\beta_\pm \geq \alpha_\pm > 0$. We give the energy spectrum and wavefunctions without details of the calculation. The case $\gamma = \kappa$ does not result in bound states whereas $\gamma = -\kappa$ does but only if $|\kappa| \geq \tau/2$ and then it has a simpler solution. Therefore, we give results for the general case but only when $\gamma \neq \pm\kappa$. Bound states are possible only if the superpotential parameters satisfy the conditions

$$\gamma\kappa < 0,\ |\gamma| > |\kappa| \geq \tau. \tag{35}$$

Following the same method used above, we obtain the following

$$\omega_n = \kappa^2 - \tau^2\left(n - |\kappa|/\tau\right)^2, \tag{36}$$

where $n = 0, 1, .., N$ and $N$ is the largest integer satisfying

$$N \leq \frac{|\kappa|}{\tau}\left[1 + \sqrt{1 + (\mu/\lambda\kappa)^2}\right]. \tag{37}$$

Equations (36) and (37) show that as $n$ increases so does $\omega_n$ which starts at $\omega_0 = 0$, reaches a maximum at or below $\kappa^2$ (when $n$ becomes the largest integer less than or equal to $|\kappa|/\tau$), then decreases towards the minimum $\omega_N \geq -(\mu/\lambda)^2$ while going through zero. Thus, $\kappa^2 \geq \omega_n \geq -(\mu/\lambda)^2$, which guarantees reality of the solution. The finite bound states energy spectrum $\{\varepsilon_n\}_{n=0}^{N}$ is obtained by substituting $\omega_n$ for $\omega$ in Eq. (14). The corresponding components of the wavefunction are

$$\phi_n^+(x) = A_n^+ \left(\cosh \tau x - 1\right)^{|\Sigma|/2} \left(\cosh \tau x + 1\right)^{-|\Delta|/2}$$
$$\times \begin{cases} {}_2F_1\left(-n, n - 2\kappa/\tau; -2n + \tfrac{1}{2}; -\sinh^2 \tfrac{\tau x}{2}\right) & , \kappa > 0 \\ (\sinh \tau x) \, {}_2F_1\left(-n+1, n+1 + 2\kappa/\tau; -2n + \tfrac{5}{2}; -\sinh^2 \tfrac{\tau x}{2}\right) & , \kappa < 0 \end{cases} \quad (38a)$$

$$\phi_n^-(x) = A_n^- \left(\cosh \tau x - 1\right)^{|\Sigma|/2} \left(\cosh \tau x + 1\right)^{-|\Delta|/2}$$
$$\times \begin{cases} (\sinh \tau x) \, {}_2F_1\left(-n+1, n+1 - 2\kappa/\tau; -2n + \tfrac{5}{2}; -\sinh^2 \tfrac{\tau x}{2}\right) & , \kappa > 0 \\ {}_2F_1\left(-n, n + 2\kappa/\tau; -2n + \tfrac{1}{2}; -\sinh^2 \tfrac{\tau x}{2}\right) & , \kappa < 0 \end{cases} \quad (38b)$$

where $\Sigma = (\gamma + \kappa)/\tau$, $\Delta = (\gamma - \kappa)/\tau$ and $n = 1, 2, .., N$. The lowest energy state is obtained from Eqs. (38) by setting $n = 0$ and requiring normalizability giving

$$\chi_0(x) = A_0^+ \left(\cosh \tau x - 1\right)^{|\Sigma|/2} \left(\cosh \tau x + 1\right)^{-|\Delta|/2} \begin{pmatrix} 1 \\ 0 \end{pmatrix}, \; \kappa > 0 \quad (39a)$$

$$\chi_0(x) = A_0^- \left(\cosh \tau x - 1\right)^{|\Sigma|/2} \left(\cosh \tau x + 1\right)^{-|\Delta|/2} \begin{pmatrix} 0 \\ 1 \end{pmatrix}, \; \kappa < 0 \quad (39b)$$

### III. CONCLUSION

In this work, we have shown that the generalized JC Hamiltonian model given by Eq. (10) is endowed with a high degree of (super-) symmetry, which we have exploited to obtain analytic solutions of various interesting examples of the model. The underlying dynamical symmetry is associated with a special four-dimensional super-algebra defined by Eq. (4). It is a $Z_2$ graded extension of su(1,1) Lie algebra which is isomorphic to the superalgebra u(1/1). A special realization of the generators of this superalgebra was employed. This realization is suitable for the description of the interaction of a bosonic field, depicted by $W(x)$, with a two-level atomic system whose effective mass (frequency) is $\mu$. The coupling strength is given by $\lambda$. Several examples with different field configurations were given. Analytic solutions (energy spectrum and state functions) were obtained. A connection between this generalized JC model and the Dirac Hamiltonian with pseudo scalar potential coupling was indicated.

### ACKNOWLEDGMENTS

This work is supported by King Fahd University of Petroleum and Minerals under project FT-2006/06. The Author is grateful to H. Bahlouli and A. Al-Hasan for fruitful discussions.